\documentclass[lettersize,journal]{IEEEtran}

\usepackage{cite}
\usepackage{graphicx}
\usepackage{amsmath,amssymb}
\usepackage{url}
\usepackage{booktabs}
\usepackage[caption=false,font=footnotesize]{subfig}
\usepackage{balance}
\usepackage{tikz}
\usepackage{eso-pic}
\AddToShipoutPictureBG*{\AtPageUpperLeft{\put(\LenToUnit{0.5\paperwidth},\LenToUnit{-0.42in}){\makebox(0,0){\footnotesize Submitted to IEEE Transactions on Visualization and Computer Graphics, 20 October 2026}}}}

\usetikzlibrary{arrows.meta, positioning}

\usepackage{xcolor}

\begin{document}

\title{TracePrism: Extending Semantic Trace Topics to Iterative, Encoding-Aware Trace Partitioning}

\author{Gennady~Andrienko,~Natalia~Andrienko,~Manuel~Resinas,~Stef~van~den~Elzen,~and~Barbara~Weber
\thanks{G. Andrienko and N. Andrienko are with Fraunhofer IAIS, Germany, and City St George's, University of London, UK.
E-mail: gennady.andrienko@iais.fraunhofer.de, natalia.andrienko@iais.fraunhofer.de}
\thanks{M. Resinas is with the University of Seville, Spain.
E-mail: resinas@us.es}
\thanks{S. van den Elzen is with Eindhoven University of Technology, The Netherlands.
E-mail: s.j.v.d.elzen@tue.nl}
\thanks{B. Weber is with the University of St.\ Gallen, Switzerland.
E-mail: barbara.weber@unisg.ch}
\thanks{This work has been submitted to the IEEE Transactions on Visualization and Computer Graphics for possible publication.}
}

\IEEEtitleabstractindextext{
\begin{abstract}
Process mining typically explains an event log with a single model, clustering, or classifier fitted to all cases, although real logs often mix cases that follow different sub-processes. Trace clustering separates sub-populations, but it fixes one notion of trace similarity for the whole analysis. We propose TracePrism, which treats process discovery as iterative partitioning with three characteristics. The trace encoding is adaptive: it is chosen for each working set of remaining traces from a library of encodings, with suggestions scored for that set. The partitioning is iterative: the analyst forms a candidate group based on a rule, a dominant topic, a cluster over topic weights, or a region of an embedding, accepts it if it is coherent in both process structure and outcome, and continues with the remaining traces. Decisions are revisable and recorded: a group can be reopened under a different working set or encoding, and the findings repository keeps both the earlier decision and its revision. We implement TracePrism as an open-source prototype and demonstrate it in usage scenarios on three real-world event logs. The scenarios also revealed behavior that we had not anticipated, such as process drift. %
\end{abstract}

\begin{IEEEkeywords}
Visual analytics, process mining.
\end{IEEEkeywords}}

\maketitle

\IEEEdisplaynontitleabstractindextext
\IEEEpeerreviewmaketitle

\section{Introduction} \label{sec:introduction}

Process mining reconstructs the behavior of a business process from event data that information systems record during execution~\cite{DBLP:books/sp/Aalst16}. An event log is a collection of traces, each an ordered sequence of events belonging to one case, for example one loan application, one hospital stay, or one traffic fine. Process discovery derives a model from the log, most often a graph of frequent transitions such as a directly-follows graph (DFG), in which nodes are activities and edges connect an activity to the activity that directly follows it~\cite{DBLP:journals/cga/MikschCSW24}.

A DFG is a single aggregate: every edge is computed over the whole case population. Real logs routinely mix cases governed by different policies, channels, or periods of the same procedure. Consider traffic fines that are paid without incident, fines for which notification fails and the case escalates to collection, and fines that take much longer and end differently again. A DFG of all three shows every activity and most transitions, but it merges three regimes into one picture that corresponds to no actual case. Filtering by variant, frequency, or case attribute helps only if the analyst already knows which partition to use. Finding that partition is the problem we address, specifically the discovery of several models of coherent sub-populations of cases instead of one model of the whole log.

Trace clustering is the usual response~\cite{TraceClustering2020,DBLP:conf/sdm/BoseA09,DBLP:journals/emisaij/JablonskiRSW18}. It represents cases under a chosen notion of similarity, partitions them into groups, and analyzes each group separately. The representation is typically selected once for the whole analysis. It then determines what similarity means: distinctions that the representation does not express cannot drive the partition. This is restrictive when sub-populations differ in different aspects of behavior. One group may be characterized by a local sequence of activities, another by timing, and another by case attributes. A representation that exposes one of them may obscure another. In addition, a group that looked plausible may become questionable once the analyst has examined it in context. This calls for revising the representation under which groups were found, and not only their number or boundaries.

The intended users are process analysts and process owners who examine the event log of a process that they want to understand or improve, with an outcome of interest in mind, such as payment, admission, or escalation. Their questions have the form: which kinds of cases exist, how do they proceed, and how do they end? An answer is a set of named sub-processes, each with a description and a known outcome profile.

Another question is when a candidate group is worth retaining. Two properties are important. The members of a group must look alike under an inspectable description of the trace, which we call \emph{structural coherence}, and they must behave alike with respect to an outcome of interest, which we call \emph{outcome coherence}. Neither is sufficient alone. Structural coherence alone can accept a group with a shared pattern but no consistent outcome, which does not help an analyst who asks what distinguishes good outcomes from bad ones. Outcome coherence alone can accept a group with similar outcomes but no process description that explains why they belong together. A system should report both to the analyst and not combine them into one score, since no fixed weighting of structural tightness against outcome purity holds across domains.
Trace clustering does not in general require outcome coherence, and outcome-oriented predictive process monitoring~\cite{ramamaneiro2023deeplearningpredictivebusiness} does not require structural coherence. Either can yield groups that are coherent in both senses, but neither uses the two as an acceptance criterion. Where no outcome label exists, any case-level partition of interest, such as a duration tier, a conformance verdict, or a cost band, can play a role instead.

Some methods iterate. Closest to our work is the interactive outcome explanation of V\"olzer et al.~\cite{DBLP:conf/icpm/VolzerZSW23}. The analyst accepts a rule over case predicates, removes the cases it covers from further search, and continues with the rest, revising the attribute vocabulary along the way. Their acceptance criteria resemble ours: an explanation must be precise with respect to the outcome and meaningful to the analyst. %
Active trace clustering~\cite{DeWeerdt2013ActiTraC}, interactive discovery~\cite{Dixit2018InteractiveDiscovery,Heijden2025interactiveDiscovery}, and process cubes~\cite{vanderAalst2013ProcessCubes} iterate over clusters, model fragments, or predefined attribute dimensions. In none of these approaches are the representation of a trace and the notion of similarity revised as the analysis proceeds. Section~\ref{sec:relatedwork} discusses these differences in more detail.

In earlier work, we proposed Semantic Trace Topics~\cite{andrienko2026semantictracetopics}, a text-inspired encoding that represents each trace as a document of tokens describing activities, local sequence, and timing. Non-negative matrix factorization (NMF) of the resulting trace-token matrix yields topics, which are interpretable, combinable components of process behavior, each tied to the traces that predominantly express it. That paper evaluated the encoding in a single pass over one working set and one encoding. It left open how to choose among encodings and the topic count, how a topic-based decomposition relates to other ways of forming groups, and what to do once an initial decomposition has been interpreted. We also build on SmartIterator~\cite{andrienko2026smartiterator}, which supports an informed choice of parameters for a single grouping step from a whole sweep of results instead of one result. TracePrism carries this principle from the group count to the working set and the encoding.

TracePrism extends trace clustering from a one-shot partition under a fixed representation to an iterative, revisable process in which the notion of trace similarity can change as different parts of the log are examined. The contributions are as follows.

\begin{enumerate}
\item \emph{Adaptive encoding.} The trace representation is chosen per working set, not once for the whole analysis. We propose a \textit{library of composable encodings} that includes token families for activities, $n$-grams, timing, and case attributes. The analyst selects encoding components guided by system-computed scores and by the analyst's knowledge of the process. %
\item \emph{Iterative partitioning.} The analyst forms a candidate group from a topic, a cluster, or a manual selection in an embedding, checks its structural and outcome coherence, commits it, and continues with the remaining traces. The commit-and-continue loop is adapted from~\cite{DBLP:conf/icpm/VolzerZSW23}. What differs is what is committed: a group formed under a trace representation that the analyst has chosen.
\item \emph{Revisability and provenance.} A committed group can be reopened and refined under a different working set or encoding. The \textit{findings repository} (as in \cite{StorySynthesis}) keeps the earlier entry, the working set and encoding under which it was formed, and how it was later revised. The analyst's growing understanding of the log is thus an explicit record.%
\end{enumerate}

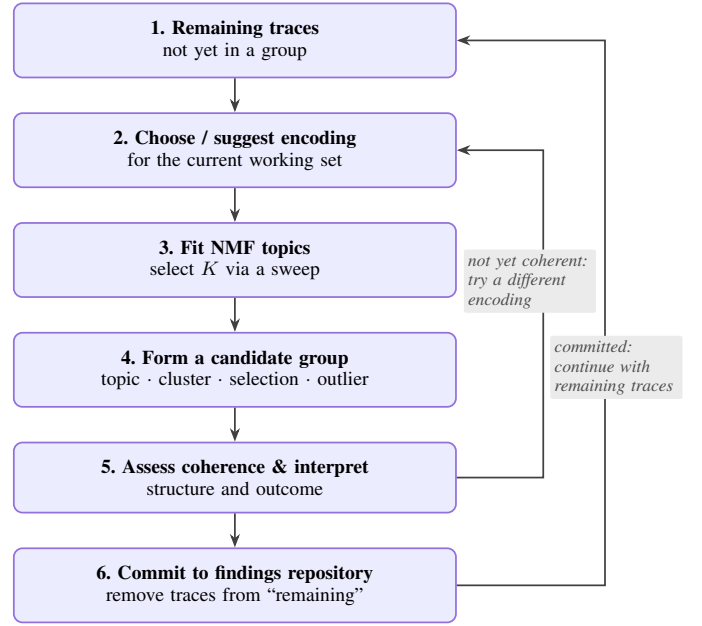
\begin{figure}[t]
\centering
\definecolor{vipraborder}{RGB}{147,112,219}
\definecolor{viprafill}{RGB}{236,236,255}
\resizebox{\columnwidth}{!}{%
\begin{tikzpicture}[
  block/.style={rectangle, rounded corners, draw=vipraborder, fill=viprafill, thick,
                text width=6.6cm, minimum height=1.0cm, align=center, inner sep=8pt, font=\small},
  lbl/.style={font=\footnotesize\itshape, color=black!70, fill=black!8,
              text opacity=1, inner sep=2pt, rounded corners=1pt},
  arrow/.style={-{Stealth[length=2.2mm]}, thick, color=black!75},
  node distance=0.55cm
]

\node[block] (n1) {\textbf{1. Remaining traces} \\ not yet in a group};
\node[block, below=of n1] (n2) {\textbf{2. Choose / suggest encoding} \\ for the current working set};
\node[block, below=of n2] (n3) {\textbf{3. Fit NMF topics} \\ select $K$ via a sweep};
\node[block, below=of n3] (n4) {\textbf{4. Form a candidate group} \\ topic $\cdot$ cluster $\cdot$ selection $\cdot$ outlier};
\node[block, below=of n4] (n5) {\textbf{5. Assess coherence \& interpret} \\ structure and outcome};
\node[block, below=of n5] (n6) {\textbf{6. Commit to findings repository} \\ remove traces from ``remaining''};

\draw[arrow] (n1) -- (n2);
\draw[arrow] (n2) -- (n3);
\draw[arrow] (n3) -- (n4);
\draw[arrow] (n4) -- (n5);
\draw[arrow] (n5) -- (n6);

\draw[arrow] (n6.east) -- ++(2.4cm,0) |- node[lbl, pos=0.2, right, align=left, text width=2cm, xshift=-.9cm]
  {committed:\\continue with\\remaining traces} (n1.east);

\draw[arrow] (n5.east) -- ++(1.4cm,0) |- node[lbl, pos=0.3, right, align=left, text width=2cm, xshift=-1.3cm]
  {not yet coherent:\\try a different\\encoding} (n2.east);

\end{tikzpicture}%
}
\caption{The iterative partitioning workflow, corresponding to tasks T1--T7 of Table~\ref{tab:tasks}. }
\label{fig:workflow}
\end{figure}

Figure~\ref{fig:workflow} summarizes the workflow. The system supports each step with visual and quantitative evidence, including a topic-stability diagram, paired coherence measures, and attribute fingerprints. Its encoding library extends~\cite{andrienko2026semantictracetopics} with additional token families and a candidate-scoring mechanism.

We implement TracePrism  as an open-source, browser-based prototype at \url{https://geoanalytics.net/TracePrism/}, and apply it to three real-world event logs from municipal traffic-fine administration, hospital sepsis care, and IT incident management. Following Isenberg et al.~\cite{isenberg2013systematic}, we evaluate the approach through usage scenarios. The claim under test is that the analysis can identify sub-populations that are coherent in structure and outcome, and record the findings. We describe a complete session on the Road Traffic Fines log in detail and compare it with the sessions on the other two logs. The sessions show that the workflow separates the logs into sub-populations that are coherent in structure and outcome, that the  findings repository remains auditable from start to end, and that the workflow applies to logs with different process structures, vocabularies, and case attributes. The auditability includes a point in the session where an earlier decision is reopened and revised on record instead of being replaced silently. 
The sessions also led to observations that we had not anticipated: groups formed from behavior alone occupy different periods of the Fines log, which points to process drift, and the outcome that resists grouping is not always the rarest one.
A formal user study is left for future work.

\section{Related Work} \label{sec:relatedwork}

\subsection{Trace encodings and trace clustering} \label{sec:rw-encodings}

Trace encoding, that is, the transformation of event sequences into feature representations, underlies clustering, anomaly detection, classification, and predictive monitoring over traces~\cite{TraceClustering2020,TAVARES2023107028,DBLP:journals/widm/RulloAS25}. \textit{Control-flow encodings} capture order and structure through $n$-grams, bags of events, or transition abstractions~\cite{DBLP:conf/sdm/BoseA09}. Song et al.\ established the profile-based pattern that many encodings still follow: a trace is a concatenation of activity, transition, performance, and case-attribute profiles, and clustering is done in the resulting space~\cite{song2009traceclustering}. \textit{Data-aware encodings} add case- and event-level attributes such as amounts, durations, or resources~\cite{DBLP:journals/emisaij/JablonskiRSW18}, and a multi-perspective extension includes sensor time series for processes~\cite{bertrand2023tropic}. \textit{Embedding-based encodings} derive numeric vectors by means of LSTMs, transformers, or autoencoders~\cite{deKoninck2018act2vec,colonna2024processminingembeddingslearning}. They often behave as black boxes and need substantial training data and tuning. In high-dimensional embedding spaces, similarity may not correspond to anything an analyst can name~\cite{CurseOfDimensionality}. We propose token families (Table~\ref{tab:encodings}) descending from the profile-based scheme. They stay discrete and interpretable, and which families participate in similarity definition is decided per working set.

In all three categories, the encoding is fixed before clustering, and the partition is reported as the result. Revisiting the notion of similarity usually means restarting the analysis. There are also approaches involving iteration. Active trace clustering extends clusters using the fitness of a model discovered from partial clusters as feedback~\cite{DeWeerdt2013ActiTraC}. The hierarchical method of Greco et al., which is closest in spirit to our round-by-round refinement, splits a cluster along the control-flow feature that best separates its members while the induced model remains too imprecise~\cite{Greco2006Discovering}. Interactive process discovery places the analyst in the model-construction loop~\cite{Dixit2018InteractiveDiscovery,Heijden2025interactiveDiscovery}, and Process Cubes slice a log along predefined attribute dimensions and discover a model per cell~\cite{vanderAalst2013ProcessCubes}. Each revisits the partition, the model, or the slice, but none revisits the trace description. The feature space is fixed before iteration begins, the splitting criterion of Greco et al.\ is chosen automatically from it, and the dimensions of a process cube are given by the log's attributes in advance. In our workflow, the analyst decides each split, the partition is discovered from learned behavioral structure, and the feature space can change between iterations. %

Surveys compare encodings mostly by predictive accuracy or clustering quality~\cite{DBLP:conf/sdm/BoseA09,DBLP:journals/emisaij/JablonskiRSW18,DBLP:journals/widm/RulloAS25}, and interpretability receives less attention. Our encoding library lies between the control-flow and data-aware families. It uses discrete, human-readable elements, applies topic modeling over the resulting vocabulary to find higher-level behavioral patterns, and makes the choice of elements an explicit, computationally-supported decision for each working set. Section~\ref{sec:casestudies} reports which encoding the scoring mechanism favors at successive rounds of analysis.

\subsection{Outcome-oriented predictive process monitoring} \label{sec:rw-predictive}

Outcome-oriented predictive process monitoring predicts case outcomes such as remaining time, next activity, compliance, or a domain-specific label from features of a partial trace. Deep architectures dominate this line of work~\cite{ramamaneiro2023deeplearningpredictivebusiness}. These methods optimize predictive accuracy. They show which features are predictive, not which traces behave alike, and their decision boundaries rarely correspond to a group an analyst could interpret and name. Interpretability is usually added afterwards through post-hoc attribution for individual predictions~\cite{galanti2020shap,rizzi2020explainability}. Such explanations are per case and per feature, and they do not show that a set of cases forms a describable behavioral pattern. Our approach considers outcome coherence together with structural coherence, instead of prediction alone.

\subsection{Subgroup and slice discovery} \label{sec:rw-subgroup}

Requiring a group to be both describable and outcome-concentrated is not specific to process mining. It is the defining requirement of subgroup discovery, which searches a space of descriptions for subsets whose target distribution deviates from the population's~\cite{wrobel1997subgroup,atzmueller2015subgroup}. Exceptional model mining generalizes the target from a label to a fitted model~\cite{duivesteijn2016emm}. Slice discovery in machine learning applies the same idea to model error and searches for underperforming, human-interpretable slices of the data~\cite{chung2019slicefinder,sagadeeva2021sliceline,eyuboglu2022domino}. Visual analytics offers interactive counterparts for subgroup exploration~\cite{subplex2022} and for auditing models across subpopulations~\cite{cabrera2019fairvis,wexler2020whatif}. Our structural and outcome coherence measures are close in spirit to the quality functions used in this literature. Unlike most of these methods, our system enables accumulating accepted groups in the findings repository instead of a final ranking.

These methods hold the descriptive language fixed: the search is over conjunctions of predicates drawn from a predetermined attribute or feature space. An exception is the interactive outcome explanation in~\cite{DBLP:conf/icpm/VolzerZSW23}. It combines rule induction over an enriched case log with separation of the cases a rule covers from the remaining search and revision of the attribute vocabulary between rounds. There are three differences from our approach. The vocabulary there is one global table revised for the analysis as a whole, while our approach enables assigning different token vocabularies to different working sets (\S\ref{sec:approach-scoring}). Their representation encodes order and timing by pre-materializing discretized attributes as distinct categorical columns, so that groups are defined through rigid conjunctions. In contrast, our approach models order and timing additively via topic distributions, enabling groups to capture soft, co-occurring token patterns that hard symbolic conjunctions fail to express. Finally, their record does not retain how each decision was reached and later revised, which our findings repository keeps as provenance (\S\ref{sec:approach-knowledge}).

\subsection{Topic modeling and NMF for process behavior} \label{sec:rw-topics}

Topic modeling finds latent themes in text collections~\cite{TM_review_2020}, and non-negative matrix factorization (NMF) is a standard way to obtain them as sparse, additive combinations of terms~\cite{lee1999nmf,luo2017NMF}. Applying it to traces requires mapping activities, transitions, or attributes to tokens, so that a trace becomes a document. Topics then correspond to co-occurring token subsets and can define process variants or behavioral patterns. We build on Semantic Trace Topics~\cite{andrienko2026semantictracetopics}, which demonstrated this scheme in a single pass: one public log, two token families, and one topic count $K$ fit over the whole log. This paper adds a larger token vocabulary with candidate scoring (\S\ref{sec:approach-encoding}--\S\ref{sec:approach-scoring}), an inspectable $K$-sweep (\S\ref{sec:approach-ksweep}), topic-based decomposition as one of three group-formation mechanisms (\S\ref{sec:approach-groups}), and a recursive process that commits, removes, or reopens groups (\S\ref{sec:approach-knowledge}).

\subsection{Visual analysis of event sequences and of topic models} \label{sec:rw-va-sequences}

Visualization research on event sequences precedes process mining. Early systems align sequences on a sentinel event~\cite{wongsuphasawat2011lifeflow,monroe2013eventflow}, extract branching patterns of frequent progressions~\cite{liu2017coreflow}, select the pattern set that compresses a collection best~\cite{chen2018sequencesynopsis}, or compare how the flow between consecutive steps diverges across cohorts~\cite{zhao2015matrixwave}. Closest to our decomposition are EventThread and its successor, which segment sequences into latent stages and cluster the stage progressions into threads~\cite{guo2018eventthread,guo2019eventthread2}. While process mining methods like the performance spectrum represent transitions as bundles of case-level trajectories—revealing micro-behaviors such as batching and overtaking—they typically fit a single holistic summary to the entire event log \cite{denisov2018spectrum}. Our framework complements this global perspective by allowing analysts to hierarchically partition traces, isolate sub-populations, and explore remaining behaviors with flexible descriptions. Integrating the two approaches is natural: performance spectra can act as per-group inspection views alongside DFGs, while stage-level segmentations can be incorporated as token families into our encoding library.

Our topic views build on designs for making topic models inspectable. Term--topic matrices with saliency-based term ordering~\cite{chuang2012termite} and coordinated relevance-ranked term views~\cite{sievert2014ldavis} represent a topic through its terms. Term ordering is a design decision. Work on comparing topic models across parameter settings~\cite{alexander2016tvcgtopic,chuang2013topicalignment} is the antecedent of our topic-stability diagram, which compares topics across values of $K$ within one corpus. 

\subsection{Multi-faceted and human-in-the-loop process exploration} \label{sec:rw-hitl}

Process exploration is an iterative, hypothesis-driven activity~\cite{SONG2008300,DBLP:books/sp/Aalst16,DBLP:journals/cga/MikschCSW24,van12Process}. Visual analytics offers a treatment of sense-making over complex, evolving data~\cite{Aigner11Visualization}, and combining it with process mining is repeatedly noted as an opportunity to go beyond purely algorithmic output~\cite{Gschwandtner17Visual,DBLP:conf/bpm/ZerbatoSW21}. The sensemaking loop alternates between foraging for evidence and building hypotheses~\cite{pirolli2005sensemaking}, which our tasks (Table~\ref{tab:tasks}) traverse. The knowledge-generation model treats the analyst's growing knowledge as part of the process~\cite{sacha2014knowledge}. Our findings repository makes this component explicit as a data structure that the analyst commits to, exports, and reopens. Analytic provenance distinguishes provenance of the data, the visualization, the interaction, and the analyst's rationale~\cite{ragan2016provenance,xu2020provenance}. The entries, action log, and reopening records of \S\ref{sec:approach-knowledge} capture interaction and state history.%

Our candidate-encoding scorer provides orienting guidance in the terms of Ceneda et al.~\cite{ceneda2017guidance}: it narrows the space of encodings to a shortlist and shows the evidence behind each suggestion, without choosing for the analyst. Exploration in practice is often centered on directly-follows graphs, which privilege control flow and constrain the questions that can be asked~\cite{Elzen2025VPA}. Coordinated multiple views and dimensionality reduction are established remedies that let the analyst inspect clusters of similar behavior against outcome or attribute distributions~\cite{cmv2007roberts,espadoto_2021,Vitale2025anomaly}. A knowledge-assisted framework treats the analyst's tacit knowledge as part of the workflow~\cite{KavaPM2026}, and human-centered visual analytics for machine learning argues that the human should be responsible for judging trustworthiness~\cite{va4hcml2022,FlowOfTrust2023}. Our structure-and-outcome judgment (\S\ref{sec:approach-coherence}) assigns this responsibility.%

Two of our earlier works are closest. SmartIterator~\cite{andrienko2026smartiterator} treats the whole sequence of grouping results across a parameter sweep as the object of inspection. We extend this sweep-based choice from a single step to a multi-stage process. Coordinated projections for multi-faceted process exploration~\cite{vandenElzen1991864} map traces into high-dimensional spaces under several encodings, reduce each to two dimensions~\cite{espadoto_2021,mcinnes2018umap}, and link the views by brushing. That approach uses boolean indicator vectors over activities, transitions, and attributes in a single pass over one subset and one pair of facets. The present paper replaces the representation with the topic-based encoding and the single pass with iterative partitioning.

\section{Analytical Tasks and Data} \label{sec:tasks}

\subsection{Task derivation} \label{sec:tasks-derivation}

\textbf{Users and needs.} TracePrism is intended for process analysts and process owners, and for the data scientists who work with them, who explore the event log of a process in order to understand or improve it~\cite{DBLP:conf/bpm/ZerbatoSW21,DBLP:books/sp/Aalst16}. They usually care about an outcome, such as whether a fine is paid, whether a patient is admitted to intensive care, or whether an incident is escalated. Their needs follow from this. They want to know what kinds of cases exist and how each kind proceeds, so a group must be describable. They want to know how each kind ends, so a group must be similar in outcome. They also want to act on the answer, for example by checking a rule, targeting an intervention, or comparing a sub-process with a reference. This requires groups that they can interpret and describe. These needs are the reason for requiring structural and outcome coherence (Section~\ref{sec:introduction}).

The analytical tasks below were derived from three sources. The first is the empirical account of exploratory process mining practices reported by Zerbato et al.~\cite{DBLP:conf/bpm/ZerbatoSW21}, which documents how analysts in practice move between orienting, searching, and interpreting, rather than following a fixed pipeline. The second is the discussion at Dagstuhl Seminar 25152 on multi-faceted visual process mining, which repeatedly identified the fixity of the trace representation as a limiting factor in current tooling. The third is the set of questions left open by our earlier single-pass formulation~\cite{andrienko2026semantictracetopics}, which map onto tasks T2 to T4 and T7 below. We phrase the tasks at the abstraction level of Brehmer and Munzner's typology~\cite{brehmer2013typology}, describing the analytical intent of each task rather than an interface action.

Table~\ref{tab:tasks} lists the seven tasks and the mechanism that supports each. The tasks are not traversed once, as in a conventional pipeline. Tasks T1--T7 form one iteration, and T7 returns to T1 over the traces that remain.

\begin{table}[t]
\centering
\caption{Analytical tasks and the mechanism supporting each. Tasks T1--T7 form one iteration; T7 returns to T1.}
\label{tab:tasks}
\footnotesize
\begin{tabular}{@{}l p{4.2cm} p{3.5cm}@{}}
\toprule
ID & Task & Supported by \\
\midrule
T1 & Assess whether the current population is behaviorally homogeneous & Data inspection; topic-stability sweep (\S\ref{sec:approach-ksweep}) \\[2pt]
T2 & Choose a trace description appropriate to the current working set & Encoding library and scoring (\S\ref{sec:approach-encoding}, \S\ref{sec:approach-scoring}) \\[2pt]
T3 & Determine an appropriate granularity of decomposition & $K$-sweep views (\S\ref{sec:approach-ksweep}) \\[2pt]
T4 & Identify a candidate group under the chosen description & Topic, cluster, or embedding selection (\S\ref{sec:approach-groups}) \\[2pt]
T5 & Judge whether a candidate is coherent structurally \emph{and} in outcome & Paired structural and outcome coherence measures (\S\ref{sec:approach-coherence}) \\[2pt]
T6 & Explain why a group differs from its parent population & Attribute fingerprints (\S\ref{sec:approach-fingerprints}) \\[2pt]
T7 & Record a decision's state and revisit it later & Repository of findings and provenance (\S\ref{sec:approach-knowledge}) \\
\bottomrule
\end{tabular}
\end{table}

\subsection{Event logs} \label{sec:tasks-data}

We exercise these tasks on three publicly available, real-world event logs. All three are widely used in the process mining literature, and they span different domains, process structures, and case-attribute vocabularies. Table~\ref{tab:datasets} summarizes their basic statistics.

\begin{table*}[t]
\centering
\caption{The three event logs used throughout this paper. Act.\ and Outc.\ denote the number of distinct activities and outcomes; Var.\ the number of distinct trace variants; Maj.\ class the share of the majority outcome class.}
\label{tab:datasets}
\footnotesize
\begin{tabular}{@{}l p{2.6cm} r r c c r c c p{2.9cm}@{}}
\toprule
Log & Domain & Cases & Events & Time span & Act. & Var. & Outc. & Maj.\ class & Case attributes \\
\midrule
Road Traffic Fines~\cite{deleoni2015RTFMlog} & municipal traffic-fine administration & 150{,}370 & 561{,}470 & 2000--2013 & 11 & 231 & 4 & 0.46 & 2 numeric, 1 categorical \\[2pt]
Sepsis Cases~\cite{mannhardt2016sepsis} & hospital emergency-department sepsis care & 1{,}050 & 15{,}214 & 2013--2015 & 16 & 846 & 5 & 0.53 & 4 numeric, 1 categorical, 11 boolean \\[2pt]
BPI Challenge 2013~\cite{steeman2013bpi2013} & IT service incident management & 7{,}553 & 65{,}532 & 2010--2012 & 13 & 1{,}511 & 3 & 0.62 & 2 numeric, 2 categorical, 1 boolean \\
\bottomrule
\end{tabular}
\end{table*}

Road Traffic Fines~\cite{deleoni2015RTFMlog} records an Italian local police fine process, from creation through notification to payment, appeal, or referral to credit collection. We adopt a published outcome specification~\cite{DBLP:conf/icpm/VolzerZSW23} that formalizes the three outcomes given in the log documentation as case predicates, namely \emph{fully paid} as a non-positive final outstanding balance, \emph{dismissed} as a dismissal code, and \emph{collected} as the occurrence of a credit-collection event, and add a fourth, \emph{unresolved}, for the cases satisfying none of the three. Its traces are short, 3.7 events on average. It is the largest of the three logs by case count, which lets us test the workflow at scale, and it is by far the most repetitive.

Sepsis Cases~\cite{mannhardt2016sepsis} records hospital emergency-department pathways for patients with suspected sepsis, with outcomes distinguishing no admission from normal-care-only, intensive-care-only, and transfers between the two. It has the richest case-attribute vocabulary of the three logs, mixing continuous vital-sign measurements (e.g., CRP, leucocytes, lactic acid) with eleven boolean clinical criteria, and it has the longest and most variable traces, 14.5 events on average and up to 185. It is the smallest log by case count, yet the most behaviorally diverse relative to its size. Since most of its attributes are boolean, it is the log most likely to stress the attribute-conjunction tokens of the encoding library.

BPI Challenge 2013~\cite{steeman2013bpi2013} records IT service incident management at Volvo IT. It contributes an escalation-oriented notion of outcome, distinct from administrative payment and from clinical admission. It tests whether the same structure-and-outcome coherence criterion transfers to a third, unrelated notion of what makes a case unremarkable or problematic.

Together, the three logs vary in case count by two orders of magnitude, in mean trace length about fourfold (3.7 to 14.5 events), in variant-to-case ratio from 0.002 to 0.81, and in case-attribute composition from three attributes to sixteen.

\subsection{Outcome definitions and preprocessing} \label{sec:tasks-outcomes}

In all three datasets, outcome labels are not recorded explicitly in the log but need to be derived from case predicates. A label that is derived partly from the activity sequence may be dependent on the control flow. We therefore state for each log how its outcome is defined and how this affects the treatment of outcome coherence.

For Road Traffic Fines, two of the four outcomes rest on evidence independent of the activity sequence: \emph{fully paid} on the final outstanding balance and \emph{dismissed} on a dismissal code. \emph{Collected} is the presence of a credit-collection event, and \emph{unresolved} is the complement of the other three, so both are partly determined by control flow. Many fines are still open at the end of the log. They are kept and form much of the \emph{unresolved} outcome. We treat outcome coherence with caution for groups dominated by \emph{collected} or \emph{unresolved}.

For Sepsis Cases, the outcome is derived from the presence of admission activities in the trace, so the label is partly determined by the same control flow that the encoding describes. The circularity is sharper here than for Road Traffic Fines. A high outcome-coherence score for a control-flow-only encoding is partly tautological, and we interpret Sepsis groups mainly through their attribute fingerprints (T6).

For BPI Challenge 2013, we define the outcome as the highest support level that an incident reaches (first, second, or third line). This is a standard quality signal in incident management reflecting how far an incident escalates through the IT hierarchy (from basic helpdesk to expert teams) before being resolved. The label is not determined by the activity sequence. It coincides, however, with a case attribute that the encoding library can tokenize. We exclude that attribute from every encoding of this log, since a topic model that uses it would reproduce the label. Resolution time and reopen count are related attributes that we do not use, and the results are specific to this choice.

Preprocessing was uniform across logs and kept minimal. No trace-length truncation was applied. Missing numeric attribute values are excluded from the corresponding fingerprint comparison instead of being imputed. Section~\ref{sec:casestudies} reports one complete session on Road Traffic Fines, with the working set, encoding, candidate group, interpretation, and commit or reopen decision at each round. The sessions on the other two logs are compared with it in Section~\ref{sec:discussion} and documented in the online supplement.

\section{Iterative Partitioning Approach} \label{sec:approach}

Figure~\ref{fig:workflow} summarizes the iterative partitioning workflow introduced in Section~\ref{sec:introduction}. The flow proceeds from the remaining traces, through choosing an encoding and fitting NMF topics for the current working set, to forming and assessing a candidate group. Two return arrows close the process rather than letting it run once and stop. If a candidate group is not yet coherent, the same working set is reopened under a different encoding. If it is coherent and interpretable, it is committed to the repository of findings, its traces are removed from the set of \emph{remaining} cases, and analysis continues with whatever traces are still left. The workflow terminates once no traces remain or no further coherent group can be found. The rest of this section describes each component of the workflow in turn.

\subsection{Trace Encoding library} \label{sec:approach-encoding}

Our earlier paper~\cite{andrienko2026semantictracetopics} proposed a vocabulary of trace tokens spanning activities, $n$-grams, lagged-follow pairs, and attribute-augmented events. %
Table~\ref{tab:encodings} lists the token families TracePrism implements, extending that vocabulary with families that make timing and case context first-class rather than incidental. These families are proposed to jointly capture three essential dimensions of business processes: control flow (local to global structure), timing and performance (delays), and case attributes. Hence, variants can be distinguished not just by what steps occurred but how and under what conditions they unfolded. While activity n-grams and lagged follows characterize sequential and proximity patterns, temporal gap tiers and relative transition times embed performance dynamics. Furthermore, attribute-augmented activities link data characteristics directly to structural execution.

The plain activity token, attribute-augmented activity, max-gap-tier, and case-attribute-tier families are order-agnostic: each reduces a trace to a set of activities or to a small number of per-trace scalar summaries, regardless of where in the sequence they occur. The remaining families are order-sensitive by construction, including immediate and short-range transitions of 2- and 3-grams and gap-augmented 2-grams, ordered but not adjacent proximity of lagged follow, and transitions labeled with the relative transition time. An encoding can combine several such families. Nothing in the workflow depends on a closed vocabulary; a new family can be easily added.

\begin{table}[t]
\centering
\caption{Trace-token families available for encoding.}
\label{tab:encodings}
\scriptsize
\begin{tabular}{@{}p{1.9cm} p{2.5cm} p{3.6cm}@{}}
\toprule
Family & Example & Captures \\
\midrule
Activity token & \texttt{act:Payment} & Activity presence \\
2-gram & \texttt{Send$\to$\allowbreak Payment} & Immediate transition \\
3-gram & \texttt{Create$\to$\allowbreak Send$\to$Pay} & Short subsequence \\
Lagged follow & \texttt{Create$\dashrightarrow$\allowbreak Payment} & Ordered proximity within a window \\
Temporal gap tier & \texttt{gap:high} & Waiting time before every event \\
Max-gap tier & \texttt{maxgap:high} & Longest wait in the trace \\
Gap-aug.\ transition & \texttt{Send$\to$Pay\allowbreak[late]} & Transition plus its own timing \\
Relative transition time & \texttt{early/\allowbreak on-time/\allowbreak late} & Timing vs.\ that transition's own norm \\
Case-attribute tier & \texttt{amount:high} & Case characteristics \\
Attribute-augmented activity & \texttt{Payment$\times$\allowbreak amount:high} & Activity--attribute conjunction \\
\bottomrule
\end{tabular}
\end{table}

Three ideas recur across the timing families of Table~\ref{tab:encodings}. A \emph{gap tier} sorts a waiting time into low, mid, or high, using cutoffs computed from the current working set rather than from the whole log. The temporal gap tier labels the wait before every event. The max-gap tier labels only the longest wait in a trace, which is why it ignores order. Gap-augmented 2-grams attach the tier of each occurrence to its 2-gram. \emph{Lagged follow} counts intervening events, not elapsed time, with a window of two by default. It thus captures a dependency such as ``eventually notified, then eventually paid''. \emph{Relative transition time} compares each occurrence's gap with the median gap of its own transition type and labels it early, on-time, or late relative to a fixed $\pm30\%$ band around that median. It thus describes the pace of a transition relative to itself.

Case-attribute tiers use cutoffs computed in the same way, over the traces that have a value. A trace missing the attribute contributes no token, because a synthetic ``missing'' tier would let incomplete traces distort the cutoffs for everyone else. Attribute-augmented activity links every token to the trace's final activity, since the logs give no other general-purpose position to privilege. Exact bin counts, window sizes, and worked examples for every family are given in the online supplement.

Together, these token families provide a sufficient feature space for identifying coherent process variants by balancing structural fidelity with performance dynamics. However, they carry inherent limitations: while short-range dependencies and local order are well-captured, processes with complex, long-range memory dependencies spanning dozens of intervening steps may require careful feature selection. Besides, the current set omits explicit resource or organizational tokens (e.g., who performed the activity). If resource allocation is a primary driver of process variant grouping, additional resource-based token families might be necessary.

\begin{figure*}[t]
\centering
\includegraphics[width=0.92\textwidth]{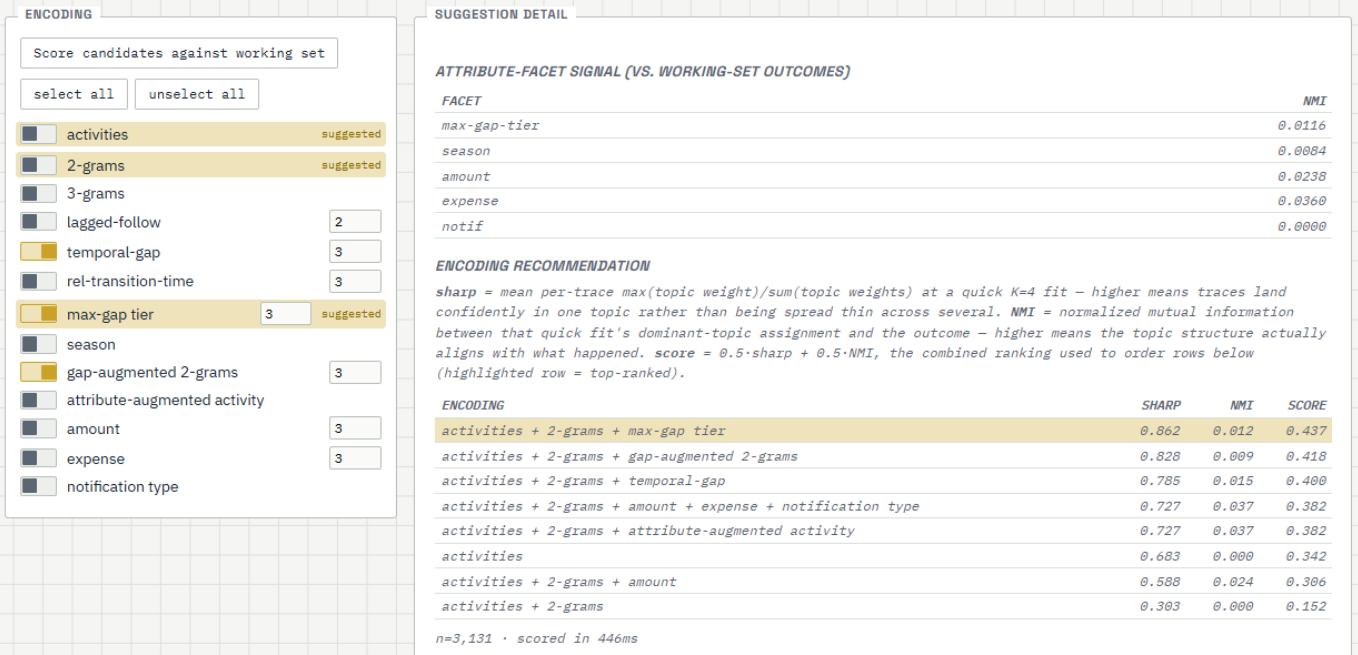}
\caption{The Encoding tab for round B of the Fines session (3{,}131 traces). Left: one switch per token family of Table~\ref{tab:encodings}; the families of the top-ranked candidate are tagged \emph{suggested}. Right: the candidate encodings ranked by score   (the highlighted row is top-ranked). The switches show the analyst's own selection, which differs from the suggestion: activities and 2-grams are off, and the temporal gap tier, the max-gap tier, and gap-augmented 2-grams are on.}
\label{fig:encoding-suggestion}
\end{figure*}

\begin{figure*}[!]
\centering
\subfloat[]{\includegraphics[width=0.45\textwidth]{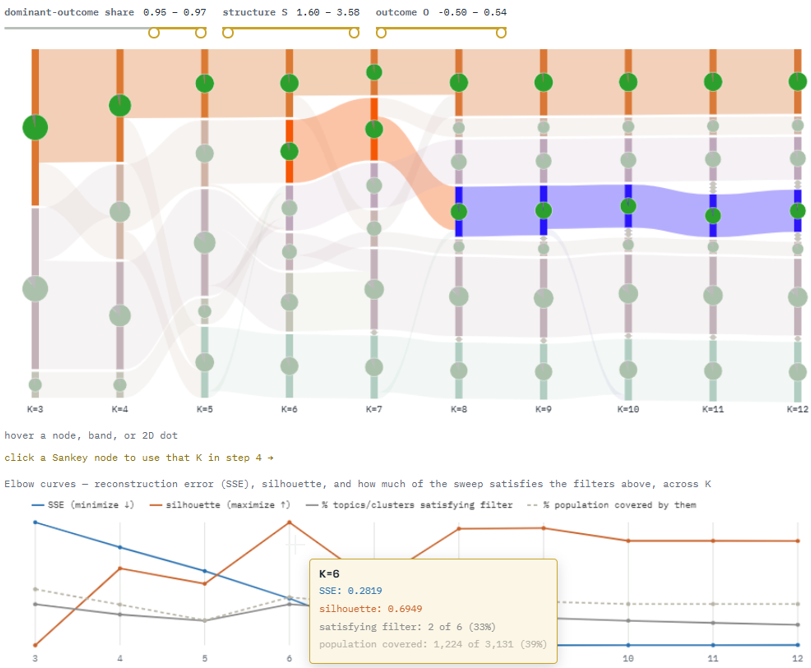}\label{fig:sys-ksweep}}
\hfil
\subfloat[]{\includegraphics[width=0.52\textwidth]{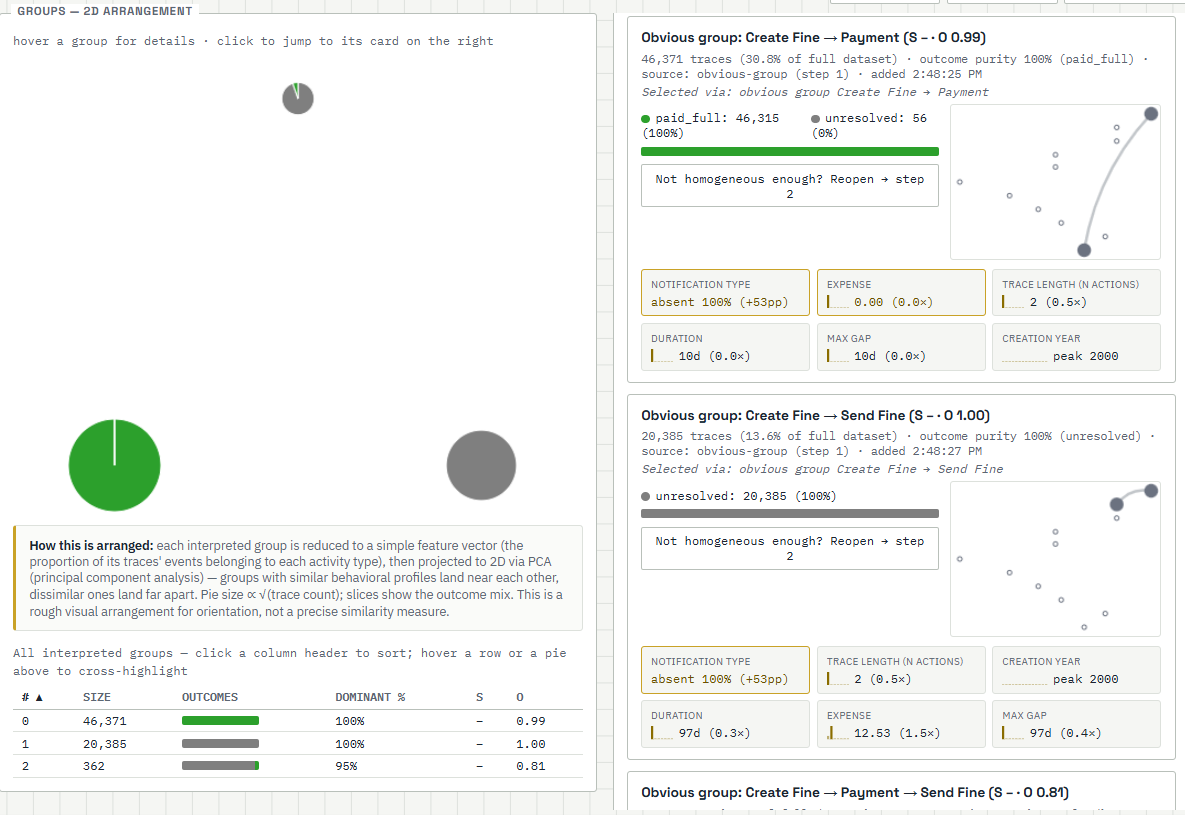}\label{fig:sys-fingerprint}}

\caption{Two of the system's mechanisms, both shown on the Fines data. (a) The K-sweep tab: a topic-stability diagram (top) tracks how topics split or merge across $K=3\ldots12$ and is linked by hover to a paired two-dimensional embedding of the same topics; the elbow, silhouette, and coverage curves below share a hover cursor with the diagram, and hovering a value of $K$ (here $K{=}6$) reports that fit's reconstruction error, silhouette score, and the share of the working set covered by the topics currently satisfying the filters.  (b) Two-dimensional arrangement of committed groups in the Knowledge tab (left), linked to two full group cards (right) with attribute-divergence information; hovering a pie in the arrangement or a row in the summary table below highlights the group in all three views: 2D plot, table, and cards.}
\label{fig:system-mechanisms}
\end{figure*}

\subsection{Encoding suggestions} \label{sec:approach-scoring}

The families of Table~\ref{tab:encodings} can be combined in many ways, and the analyst may set any combination by hand. To help the analyst start, TracePrism scores the candidate encodings on the current working set and ranks them as suggestions.  The analyst may adopt a suggestion, adjust it, or ignore, and the choice may depend on what the analyst knows about the process and about what earlier rounds have covered. For example, where timing is expected to matter, the gap families are natural candidates, and after committing groups based on common 2-gram patterns, 3-grams are a natural next step. Figure~\ref{fig:encoding-suggestion} shows the Encoding tab for round B of the Fines session, in which the analyst's selection differs from the top suggestion.

For each candidate, TracePrism fits a quick NMF model with $K{=}4$ topics and computes two indicators. Sharpness, $\mathit{sharp}$, is the mean over traces of each trace's largest topic weight divided by the sum of its topic weights; it is high when traces land clearly in one topic. Normalized mutual information, $\mathit{NMI} = 2\,\mathrm{MI}(T;Y)/(H(T)+H(Y))$, is computed between the dominant-topic assignment $T$ and the outcome label $Y$; it is high when the topic structure aligns with what happened to the cases. The suggestion score is their plain average, $\mathit{score} = 0.5\cdot\mathit{sharp} + 0.5\cdot\mathit{NMI}$, a neutral default that only orders the list. Both indicators are also shown separately, so the analyst can see whether a high score stems from separation, from outcome alignment, or both.

We fix $K{=}4$ for three reasons. A common $K$ gives all candidates the same number of topics, which keeps their indicators comparable. A small $K$ keeps each fit cheap enough to repeat whenever the working set changes. And $K{=}4$ is close to the number of outcome classes of our three logs (four, five, and three), so the topic count does not limit how well the dominant topic can align with the outcome. The value is a coarse probe of whether an encoding separates the traces into a few clear groups. It is not an estimate of the right topic count, which the $K$-sweep examines next (\S\ref{sec:approach-ksweep}). The score is a recommendation and not a verdict; the judgment on a group remains with the analyst (\S\ref{sec:approach-coherence}).

\subsection{NMF topic fitting and the K-sweep} \label{sec:approach-ksweep}

Once an encoding is chosen, traces become documents over the resulting token vocabulary. A token's raw frequency within a trace is counted directly, so an activity or transition that recurs several times contributes more weight before normalization, and each frequency is scaled by the standard smoothed inverse document frequency $\mathrm{idf}(t) = \ln\frac{1+n}{1+\mathrm{df}(t)} + 1$, where $n$ is the number of traces in the current working set and $\mathrm{df}(t)$ the number of them containing token $t$. Every trace's resulting weight vector is then normalized to unit Euclidean length, which is what keeps a long trace and a short one on a comparable scale rather than letting trace length alone drive the fit. NMF is applied to the resulting trace--token matrix, as in~\cite{andrienko2026semantictracetopics}, to obtain a topic--term matrix and a trace--topic matrix. Rather than fixing the topic count $K$ once, TracePrism sweeps $K$ over an analyst-chosen range and presents  coordinated views of the result (Figure~\ref{fig:system-mechanisms}): a topic-stability diagram, a Sankey-style layout linking each topic at one $K$ to the topics it splits from or merges into at adjacent values of $K$; a reconstruction-error (elbow) curve; and a silhouette curve~\cite{rousseeuw1987silhouettes}, computed over a capped sample of traces for tractability at scale. Topics that remain stable across a run of adjacent $K$ values are a visual sign that they capture something robust to the exact topic count. Topics that fragment or merge erratically signal that $K$ is not yet well matched to the working set. The elbow and silhouette curves support this reading quantitatively, but do not replace it: a stable, interpretable topic at a locally poor silhouette value is still preferred over an unstable one at a locally better one.

\subsection{Three group-formation mechanisms} \label{sec:approach-groups}

Given a fitted trace--topic matrix, TracePrism supports three interchangeable ways of forming a candidate group. They differ in how much of the topic-weight vector they use and in how much the analyst contributes. Table~\ref{tab:groups} lists their strengths and weaknesses.

\emph{Dominant topic.} The group consists of the traces for which one topic has the largest weight. This is the simplest mechanism, and the group is described by the top terms and the process graph of its topic. The assignment uses only which weight is largest. It ignores the size of that weight and its lead over the second topic, so a trace that splits its weight evenly between two topics is assigned as firmly as a trace that belongs to one topic.

\emph{$k$-means over topic weights.} The group is a $k$-means cluster over the full topic-weight vectors. Since the whole vector is used, a cluster can capture a coherent mixture of topics that no single topic isolates. The vectors have $K$ dimensions (at most 21 in our sessions), unlike the indicator vectors of~\cite{vandenElzen1991864}, which have one dimension per activity, transition, and attribute. Distances are less affected by the loss of contrast in high dimensions~\cite{CurseOfDimensionality}, and clustering is cheap. The number of clusters $N$ is chosen with a sweep and a silhouette curve, as $K$ is. A cluster has no topic of its own, so it is harder to interpret. The analyst reads it through the topics that its members weight most, its process graph, and its fingerprint.

\emph{Manual selection in an embedding.} The group is a region that the analyst brushes in a two-dimensional UMAP embedding~\cite{mcinnes2018umap} of the topic-weight vectors. Like clustering, it uses the composition of the weights, but, unlike in clustering, the analyst can follow irregular shapes and partial overlaps between topics, and can apply knowledge of the process to decide where a group ends. However, the embedding is a projection and has artifacts. Distances and shapes are distorted, and a region that looks compact may not be coherent in the full space. %

\begin{table}[h]
\centering
\caption{Group-formation mechanisms compared.}
\label{tab:groups}
\footnotesize
\setlength{\tabcolsep}{3pt}
\begin{tabular}{@{}p{1.45cm} p{3.2cm} p{3.4cm}@{}}
\toprule
Mechanism & Advantages & Disadvantages \\
\midrule
Dominant topic & Simple; described by the terms and graph of its topic & Ignores the size of the dominant weight and its lead over the next topic; ignores the composition of topics \\[2pt]
$k$-means over topic weights & Reflects the composition of topics; $N$ chosen with a sweep & Harder to interpret: no topic of its own \\[2pt]
UMAP selection & Reflects the composition of topics; any region shape; uses the analyst's judgment & Projection artifacts; manual effort \\
\bottomrule
\end{tabular}
\end{table}

We keep the three mechanisms interchangeable, rather than combining them into one procedure, because they encode different notions of what makes traces belong together: the leading topic, the composition of topics, or a region that the analyst sees. Forcing a single definition would reintroduce the commitment that the rest of the workflow is designed to avoid. Which notion is appropriate depends on the working set and the question, and it can change from one iteration to the next.

\subsection{Structure-and-outcome coherence} \label{sec:approach-coherence}

A candidate group $G$ drawn from a population $D$, which is a working set or the whole log, is accepted only if it is coherent in both the structural and the outcome sense introduced in Section~\ref{sec:introduction}. Denoting the topic-weight vector of trace $i$ under the current encoding as $w_i \in \mathbb{R}^K$ ($\mathbb{R}^K$ is the topic-weight space with $K$ dimensions), we define structural coherence as the ratio $S(G) = s(G)/s(D)$, where $s(\cdot)$ is the mean pairwise cosine similarity of the topic-weight vectors of a set of traces. A value above 1 means that the members of $G$ are more alike than a typical pair of traces in $D$. Groups whose members sit close together in the topic-weight space score highly, regardless of which of the three mechanisms of Section~\ref{sec:approach-groups} produced them. Because the topic-weight space of $D$ is refit at every round, $S(G)$ is comparable across candidate groups drawn from the same round's fit, not across rounds with a different $K$ or a different working set. 

Writing $p_D$ and $p_G$ for the outcome-label distributions over $D$ and $G$, we define outcome coherence as $O(G) = 1 - H(p_G)/H(p_D)$, where $H(\cdot)$ is Shannon entropy. $O(G)$ approaches 1 when $G$'s outcomes concentrate on a single label, and approaches 0 when the outcome mix in $G$ is close to the one in the population $D$, that is, becomes no more informative than chance. It can be slightly negative for a small group whose outcomes are, by chance, more spread out than the population's, and it is undefined when $D$ has a single outcome. Because it depends only on entropy, $O(G)$ cannot distinguish a group that isolates the population's already-dominant outcome from one that inverts it: a population split $(0.9,0.1)$ and a group split $(0.1,0.9)$ have equal entropy and so the same $O(G)$, even though the group's dominant outcome is the population's rarest. We rely on the attribute fingerprint and the analyst's own reading of the outcome mix, not on $O(G)$ alone, to catch this case.

Both quantities are reported to the analyst rather than thresholded automatically. A group with high $S(G)$ but low $O(G)$ is structurally clean but behaviorally uninteresting. A group with high $O(G)$ but low $S(G)$ is behaviorally striking but not yet describable. The analyst judges whether a particular combination is worth committing. This cannot be done automatically: no fixed weighting of $S(G)$ against $O(G)$ is applicable across domains, so the trade-off is presented as a two-dimensional judgment (task T5) rather than reduced to a single score.

\subsection{Attribute fingerprints} \label{sec:approach-fingerprints}

\begin{figure}[t]
\centering
\includegraphics[width=\columnwidth]{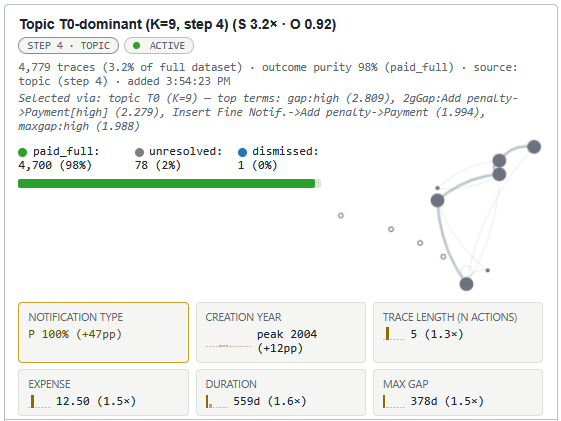}
\caption{Information card of the committed group T0 of round D ($K{=}9$, 4{,}779 traces) with its attribute fingerprint. The header gives $S(G)=3.2\times$ and $O(G)=0.92$, and the bar shows the outcome mix (98\% \emph{paid\_full}). Each tile compares one attribute with the source population. Numeric tiles show the group mean and its ratio to the population mean, but the highlighting is decided by the standardized effect size. Categorical tiles show the category whose share differs most from the population: here the value P of notification type makes up 100\% of the group, 47 percentage points above the population. The creation-year tile shows the year with the largest share difference; at +12 points (rounded) it stays just below the threshold. Only the highlighted tile (notification type) clears the threshold.}
\label{fig:fingerprint-card}
\end{figure}

To help the analyst interpret why a candidate group is or is not coherent in outcome, every group is automatically compared against its source population on every case attribute for which there is enough data. Numeric attributes are compared using a standardized effect size, which is the difference between the group's and the population's mean, scaled by the population's own spread, so that an outlying value means the same thing across attributes with very different natural variability. Categorical and boolean attributes are compared as a percentage-point difference in category share. If the data include the case creation time, its distribution over a group is compared as one of the facets, since a temporal concentration usually reflects a policy change or a seasonal effect rather than an intrinsic case property. To avoid clutter, the system only displays attributes that deviate significantly from the baseline: specifically, numeric variables with an effect size of at least 1.2, or categorical variables with a share difference of at least 12 percentage points. Groups with fewer than eight traces get no fingerprint, so that a handful of incidentally unusual values in a small group do not appear as a pattern. The full comparison remains available on demand once a group is committed. Figure~\ref{fig:fingerprint-card} shows the fingerprint of topic T0 from round D of the Fines session (Section~\ref{sec:casestudies}). In the compact views of the system, only the highlighted attributes are shown.

\subsection{Findings repository, reopening, and provenance} \label{sec:approach-knowledge}

A committed group becomes an entry in the findings repository, recording its traces, outcome and attribute profile, source mechanism, encoding, and timestamp. Committing a group temporarily removes its traces from the \emph{remaining} pool of data. The proportion of all traces currently locked in committed groups defines the repository’s coverage, while the remaining unassigned traces form the residue. Reopening a group does not delete it; instead, the entry is archived in history while its traces are released back into the remaining pool for further analysis. Because reopening releases cases without erasing history, a trace can belong to multiple entries: an archived past group and a newly committed one. The repository tracks both the current committed group for every trace and its complete history. This provides interaction and analysis-state provenance: recording which groups were created, modified, or reopened, and under what conditions. This history is exported alongside the findings to preserve the full analytical trajectory rather than just the final state.

\section{System} \label{sec:system}

We implement the workflow of Section~\ref{sec:approach} as TracePrism, a single-page, browser-based application with no server-side component. An event log is loaded, held, and processed entirely client-side, so no data leaves the analyst's machine. The two computationally heavy steps, the repeated NMF fits of the $K$-sweep (\S\ref{sec:approach-ksweep}) and the UMAP embedding used for manual selection (\S\ref{sec:approach-groups}), run in dedicated Web Workers, off the browser's main thread, so the interface stays responsive while a sweep or an embedding is computed. $k$-means clustering over topic weights is comparatively cheap and runs on the main thread alongside the interface itself. Section~\ref{sec:discussion} reports what each step costs in practice. The system is open source. The implementation and the three event logs of Table~\ref{tab:datasets} are available in the online supplement at \url{https://geoanalytics.net/TracePrism/}.

The interface is organized into seven tabs, Data, Encoding, K-sweep, Topics, Trace groups, Knowledge, and Log, which together cover the six numbered stages of Figure~\ref{fig:workflow} and add a session log; every task T1--T7 of Table~\ref{tab:tasks} is supported by at least one tab. The tabs are not seven independent screens; they are coordinated views over a single shared trace-and-encoding state, so a selection, a commit, or a toggle in one tab is reflected in all the others without an explicit refresh. This coordination is what lets the workflow of Figure~\ref{fig:workflow} be driven interactively rather than scripted. A full tab-by-tab mapping to workflow stage, task, and where each computation runs is given in the online supplement; the rest of this section summarizes each tab's role.

\begin{figure}[t]
\centering
\includegraphics[width=\columnwidth]{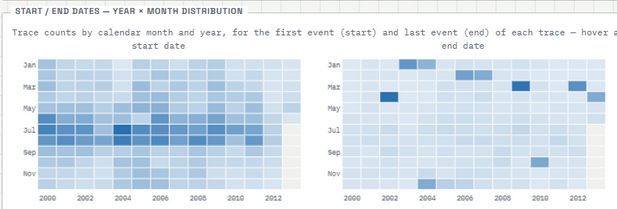}
\caption{Trace counts by calendar month and year, for each trace's first event (start, left) and last event (end, right), from the Data tab.}
\label{fig:casestudy-dates}
\end{figure}

\textbf{Data} supports T1 and, through rule-based groups, T4. Sortable distribution plots over activities, durations, and case attributes, including a year-by-month heatmap of each trace's start and end date (Figure~\ref{fig:casestudy-dates}), support an overview-then-detail viewing of the working set. Groups meeting a size and outcome purity threshold are detected automatically and committed with one click.

\textbf{Encoding} supports T2. Every token family of Table~\ref{tab:encodings} is a switch, and the suggestions are listed with their sharpness, normalized mutual information (NMI), and aggregated score. The analyst reads the suggestions, selects the encoding, and either proceeds to the K-sweep tab or decides to look at a different working set.

\textbf{K-sweep} supports T1 and T3. The sweep range sets how many NMF fits are run, one for each $K$ in the range. The fits run one after another in a worker, so the interface stays responsive, and the results appear in linked views (Figure~\ref{fig:sys-ksweep}), including a coverage curve: the share of the working set that passes the coherence filter. Three range sliders, on dominant-outcome share, $S(G)$, and $O(G)$, then highlight the topics whose values fall within the selected ranges. Moving a slider updates the highlighting without new computation. The sliders help to narrow the range of $K$ to a few candidates for inspection. They do not decide which topics are committed.

\begin{figure*}[t]
\centering
\includegraphics[width=.85\textwidth]{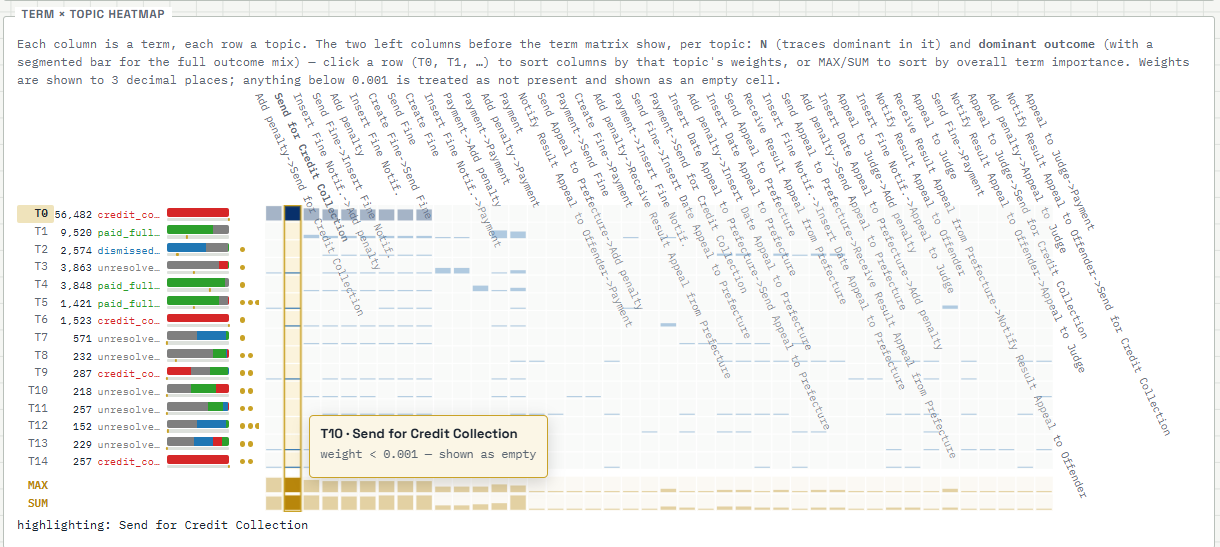}
\caption{The term--topic weight matrix underlying round C's $K{=}15$ fit, one row per topic and one column per vocabulary term, with cell shading giving each term's NMF weight in that topic. The two columns to the left of the matrix summarize each topic at a glance: $N$, the trace count for which the topic is dominant, and its dominant-outcome bar, showing the full outcome mix rather than only the majority share; each dot beside the bar marks an attribute whose fingerprint (\S\ref{sec:approach-fingerprints}) clears the divergence threshold for that topic. Columns can be sorted by a topic's own weights (here T0) or by importance across the whole fit (MAX and SUM rows). Weights below 0.001 are treated as absent and shown as an empty cell.}
\label{fig:casestudy-heatmap}
\end{figure*}

\begin{figure}[t]
\centering
\includegraphics[width=\columnwidth]{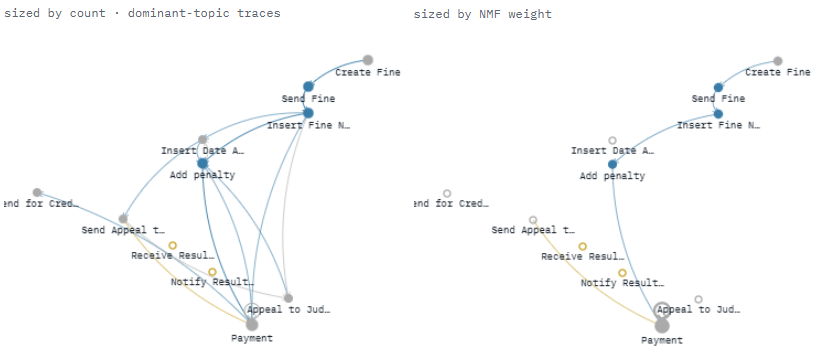}
\caption{The directly-follows graph of a single committed topic (T4, $K{=}15$, 3{,}848 traces), drawn twice to separate two notions of importance that a single sizing would conflate. The left panel sizes elements by how often they occur among T4's dominant traces; the right panel sizes them by the weight NMF assigns, which can be nonzero even for a transition absent here, if the term is shared with other topics. Color compares each element's weight and count, each divided by T4's maximum: gold marks the 15\% of elements where the relative weight exceeds the relative count the most (distinctive of T4), blue the 15\% where the relative count exceeds the relative weight the most (frequent but not distinctive), and gray the rest.}
\label{fig:casestudy-dualdfg}
\end{figure}

\begin{figure}[t]
\centering
\includegraphics[width=\columnwidth]{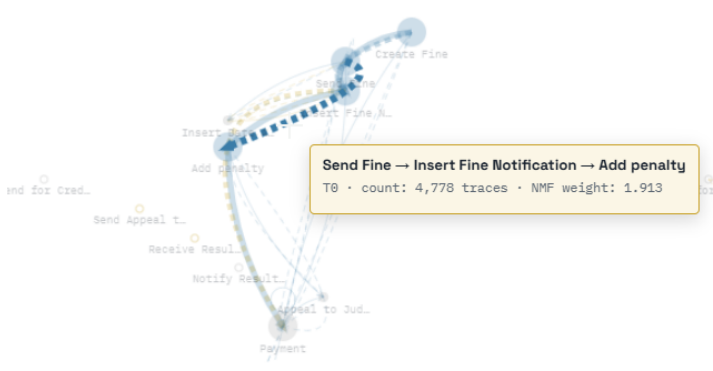}
\caption{Hovering over an element of a topic's graph reports its trace count and its NMF weight. The highlighted dashed edge is the triple \emph{Send Fine$\to$Insert Fine Notification$\to$Add penalty} of topic T0 in round D (4{,}779 traces). It is followed by 4{,}778 of the topic's traces and has a weight of 1.913.}
\label{fig:edge-tooltips}
\end{figure}

\textbf{Topics} supports T4, T5, and T6. In response to selecting a topic from the stability diagram, a small-multiples overview, or a term--topic heatmap (Figure~\ref{fig:casestudy-heatmap}), the system shows its directly-follows graph, $S(G)$/$O(G)$ value, and attribute fingerprint. 
The graph is drawn in two variants, with nodes and links sized by either raw count or NMF weight. Color-coding shows how far the two variants diverge (Figure~\ref{fig:casestudy-dualdfg}). This allows separation of topic-defining behavior from what just occurs frequently among traces containing this behavior (T6). Hovering over an element reports its trace count and NMF weight (Figure~\ref{fig:edge-tooltips}).
An encoding with 3-grams adds a dashed edge per salient triple. Each fingerprint tile (which is shown only if assessed to be informative) reports a standardized effect size, a percentage-point difference, or a creation-date facet as appropriate (\S\ref{sec:approach-fingerprints}); worked examples are given in the online supplement.

\textbf{Trace groups} supports T4 and T5, implementing the two group-formation mechanisms besides the dominant topic. An $N$-sweep slider previews $k$-means cluster counts $N{=}3\ldots20$ against a live silhouette curve; a brushable two-dimensional UMAP embedding supports manual selection where neither a topic nor a cluster isolates a coherent subset cleanly. The embedding parameters (neighborhood size, minimum distance, spread, and seed) are exposed and held fixed, so a brushed region can be reproduced. Both mechanisms report $S(G)$ and $O(G)$ live as the selection changes.%

\textbf{Knowledge} tab supports T5 and T7 by displaying the findings repository and enabling operations on it. Committed groups are represented by pie glyphs in a 2D arrangement obtained using PCA, where similar entries tend to be positioned near each other, and also in a sortable summary table (Figure~\ref{fig:sys-fingerprint}). For each group, there is a card with detailed information (Figure \ref{fig:fingerprint-card}). A group can be reopened, which restores the exact working set and encoding under which the group was formed. Every entry exports as CSV with its provenance; this is how the tables of Section~\ref{sec:casestudies} were obtained.

\textbf{Log} is a plain, timestamped, unfiltered list of every session action. It is not tied to any specific task. Its purpose is to make provenance (T7) checkable after the fact: every number and figure in Section~\ref{sec:casestudies} is drawn directly from a session log and findings repository.

\section{Case Studies} \label{sec:casestudies}

This section reports one complete iterative partitioning session on Road Traffic Fines, the largest of the three logs, chosen to demonstrate the workflow at a level of complexity where a single global model fails to capture underlying variations. The sessions on Sepsis Cases and BPI Challenge 2013 are documented step by step in the online supplement and compared with this one in Section~\ref{sec:discussion}. The Fines session touches every task of Table~\ref{tab:tasks}.

\textbf{Obvious groups (T1, T4).} Before any modeling, three exact-variant groups met the size and purity thresholds and were committed directly from the Data tab: \emph{Create Fine$\to$Payment} (100\% \emph{paid\_full}), \emph{Create Fine$\to$Send Fine} (100\% \emph{unresolved}), and \emph{Create Fine$\to$Payment$\to$Send Fine} (95\% \emph{unresolved}). Together they cover close to half the log.

\textbf{Encoding choice (T2).} A fourth, near-obvious variant, \emph{Create Fine$\to$Send Fine$\to$Payment}, did not pass the purity threshold and went to modeling. Table~\ref{tab:fines-final} lists for each round the encoding that the scorer ranked first and the one that the analyst applied. The analyst's choices drew on knowledge of the process. In round B all traces share the same activity sequence, so activities and 2-grams carry no signal. The analyst switched them off in favor of three timing families (Figure~\ref{fig:encoding-suggestion}). The date distributions (Figure~\ref{fig:casestudy-dates}) show that the start dates spread continuously while the end dates cluster in monthly bands, which suggests batch processing at closing. Round C used the top-ranked activities and 2-grams. By round D the obvious 2-gram patterns belonged to committed groups, so the analyst added 3-grams to capture longer sequences and kept the timing families. Rounds E and F reuse the encoding of round D.

\begin{table*}[t]
\centering
\caption{The Fines session by round. Top-ranked is the encoding that the scorer ranked first; applied is the analyst's selection. Coverage is the share of the 150{,}370 traces that belong to an accepted group (running total, net of the round D reopening, Table~\ref{tab:fines-reopen}); the remaining traces form the residue. Purity is the dominant outcome's share, min--max over the round's committed groups.}
\label{tab:fines-final}
\scriptsize
\setlength{\tabcolsep}{3pt}
\begin{tabular}{@{}c r p{2.6cm} p{3.9cm} c l r c r@{}}
\toprule
Round & Working set & Top-ranked encoding & Applied encoding & $K$ & Committed (groups, cumulative) & Traces & Purity & Coverage \\
\midrule
A & 150{,}370 & -- & obvious variants (no model) & -- & 3 (3) & 67{,}118 & 95--100\% & 44.6\% \\
B & 3{,}131 & act.\ + 2-grams + max-gap tier & temporal gap + max-gap tier + gap-aug.\ 2-grams & 6 & 3 topics (6) & 1{,}818 & 94--97\% & 45.8\% \\
C & 81{,}434 & act.\ + 2-grams & act.\ + 2-grams & 15 & 4 topics (10) & 62{,}110 & 93--100\% & 87.1\% \\
D & 19{,}324 & act.\ + 2-grams & timing families + 3-grams & 9 & 4 topics$^\dagger$ (14) & 10{,}561 & 86--98\% & 94.2\% \\
E & 8{,}763 & not rescored & as in D & 20 & 5 topics (19) & 3{,}742 & 90--96\% & 96.7\% \\
F & 5{,}021 & not rescored & as in D & 21$^\ast$ & 3 clusters (22) & 915 & 72--84\% & 97.3\% \\
\midrule
\multicolumn{6}{@{}l}{Total committed / residue} & 146{,}264 / 4{,}106 & -- & 97.3\% / 2.7\% \\
\bottomrule
\end{tabular}
\\[3pt]
{\raggedright\scriptsize $^\dagger$A fifth topic from round D (T8) was committed and then reopened before round E (Table~\ref{tab:fines-reopen}). $^\ast$Topics were fit only to supply topic-weight vectors for clustering and the UMAP view; the committed groups are $k$-means clusters ($N{=}12$) over those weights.\par}
\end{table*}

\textbf{The headline group (T4--T6).} Round C ($K{=}15$) produced 15 topics, four of which passed the coherence filter. Figure~\ref{fig:casestudy-heatmap} shows the fit as a term--topic matrix. The largest, T0, holds 56{,}482 traces, all with outcome \emph{credit\_collection}, and has a nearly linear control-flow signature. It carries no attribute fingerprint: no case attribute differs enough from the population to be flagged (\S\ref{sec:approach-fingerprints}). The group is thus coherent through control flow alone, although its outcome label is itself defined by a credit-collection event, so its outcome coherence is partly by construction.

\textbf{Reopening and provenance (T7).} Round D produced a fifth candidate, T8, which was committed although its outcome coherence was only moderate (Table~\ref{tab:fines-reopen}). The analyst later reopened it from the Knowledge tab, which returned its traces to the remaining set and made them part of round E's working set. After the refit, most of them landed together in a new topic with higher purity, and the rest fell back into the residue. The superseded entry stays on record as it was, so the original decision and its revision both remain inspectable.

\begin{table}[t]
\centering
\caption{The reopened entry (T8, $K{=}9$) and the fate of its 860 traces after reopening.}
\label{tab:fines-reopen}
\scriptsize
\begin{tabular}{@{}p{3.8cm} r c l@{}}
\toprule
State & $n$ & $O(G)$ & Dominant outcome \\
\midrule
Committed, round D (superseded) & 860 & 0.48 & credit\_collection, 83.5\% \\[2pt]
$\hookrightarrow$ recaptured, round E (T3, $K{=}20$) & 528 & 0.87 & credit\_collection, 96.2\% \\[2pt]
$\hookrightarrow$ back to residue & 332 & -- & mixed \\
\bottomrule
\end{tabular}
\end{table}

\textbf{Diminishing returns and stopping (T1).} Round F, on the residue left after round E, is the weakest round. A cluster-count sweep and two UMAP views yielded only three clusters that passed the coherence filter, with 72--84\% purity. We include it as the concrete case of the termination condition in Figure~\ref{fig:workflow}: it shows what \emph{no further coherent group can be found} looks like in practice.

\textbf{Final tally and residue.} Table~\ref{tab:fines-final} lists the groups committed in each round and the resulting coverage. Coverage rises with every round, and the figures are net of the group that was reopened in round D (Table~\ref{tab:fines-reopen}). The 22 active groups cover 97.3\% of the log (146{,}264 traces), and 4{,}106 traces remain in the residue. The residue is not a random sample of the log. It consists mainly of two outcomes. \emph{Dismissed} traces are the hardest to group: 84.3\% of all dismissed traces (1{,}737 of 2{,}060) remain in the residue, and they make up 42.3\% of it. \emph{Unresolved} traces are mostly grouped, since only 5.7\% of them remain (1{,}663 of 29{,}010), but they are numerous, so they still make up 40.5\% of the residue. Residual traces are also longer (mean 6.1 events against 3.7 overall), as expected if topic weights of long traces scatter across several topics. Section~\ref{sec:discussion} compares this pattern with the other two logs.

\section{Discussion} \label{sec:discussion}

\begin{table}[t]
\centering
\caption{The three case-study sessions compared. }
\label{tab:crosslog}
\scriptsize
\begin{tabular}{@{}l r c c p{3.7cm}@{}}
\toprule
Log & Traces & Rounds & Coverage & Hardest outcome to cover \\
\midrule
Fines & 150{,}370 & 6 & 97.3\% & dismissed (84.3\% in residue) \\[2pt]
Sepsis & 1{,}050 & 5 & 87.2\% & ICU-escalation outcomes jointly; IC\_only worst (90.0\%) \\[2pt]
BPI2013 & 7{,}553 & 3 & 54.3\% & reached\_line\_2 (72.6\% in residue) \\
\bottomrule
\end{tabular}
\end{table}

\textbf{Analysis outcomes.}
The outcome of a TracePrism session is a comprehensive record of findings: what was discovered, under which working set and encoding, and what was subsequently revised. Across three real event logs (traffic-fine administration, sepsis care, and IT incident management), the accepted groups covered 97\%, 87\%, and 54\% of the traces, respectively. Because the un-grouped outcomes varied from rare to mid-frequency across these logs, data rarity alone does not determine how difficult a pattern is to capture.

Crucially, these coverage percentages are secondary to the analytical record itself. For instance, one session initially accepted a group of 860 traces, 83.5\% of which ended in credit collection, only for the analyst to later reopen it. Under a refined working set, 528 of those traces formed a tighter group with 96.2\% purity, while the remaining 332 returned to the pool; both the initial decision and its subsequent correction remain part of the permanent record. This detailed record also enables unplanned secondary analyses. When post-hoc checks examined the time spans of groups formed entirely without calendar information, distinct temporal clustering emerged: some groups dominated the early years of the log while others appeared only in later ones. Because the base of findings explicitly tracks trace membership, this verification required no new modeling, directly revealing process drift and behavioral evolution over time.

\textbf{Generality across domains.} Table~\ref{tab:crosslog} summarizes all three sessions. Coverage is the share of traces that belong to an accepted group, under each session's own thresholds. Rounds counts committing rounds only: the obvious groups of Round A plus every modeling round that produced at least one group. The last column gives the outcome with the largest share of its traces left in the residue.

The Fines session is detailed in Section~\ref{sec:casestudies}, and the other two are documented in the online supplement. Coverage drops from Fines (97.3\%) to Sepsis (87.2\%) to BPI2013 (54.3\%). The three logs also disagree on which outcome is hardest to cover. For Fines it is \emph{dismissed}, a genuinely rare outcome (1.4\% of the log). For Sepsis it is the three ICU-escalation outcomes jointly, none of which ever produced a coherent topic or cluster, with \emph{IC\_only} the worst at 90.0\% of its instances left in the residue. For BPI2013 it is \emph{reached\_line\_2}, which is neither the rarest nor the most common of three outcomes; the rarest, \emph{reached\_line\_3}, was isolated almost immediately, at perfect purity, in the first topic round.

Read together, these three sessions led to a finding that we did not expect: rare outcomes are not always the hardest to group. Whether an outcome resists partitioning depends on whether it corresponds to a distinct behavioral regime under some encoding, not on its base rate alone, and exposing that correspondence, rather than assuming it, is the purpose of the workflow's structure-and-outcome coherence criterion.

\textbf{Encoding choice across sessions.} Table~\ref{tab:crosslog}'s Rounds column counts committing rounds, not encoding changes, so the latter is worth stating directly. We distinguish the encoding that the scorer ranks first from the one that the analyst applies. In the Fines session the ranking changes once, in round B. The applied encoding changes in each of the first three modeling rounds: timing families only in round B, activities and 2-grams in round C, and a selection with timing and 3-gram families in round D, which rounds E and F reuse. These departures from the ranking reflect the analyst's knowledge of the process. Sepsis holds one encoding, activities, 2-grams, and attribute-augmented activity, across all four of its rounds. BPI2013 evaluates two different encodings: one for the initial "easy majority" of cases and another for the harder remaining cases. It drops the case-attribute tokens used for the majority group and switches to a gap-augmented control-flow encoding for the final clustering round. Overall, the encoding changes only when the nature of the data shifts (such as moving from a majority group to a harder residue) or when the analyst's domain knowledge requires it, rather than changing in every round.

\textbf{What the Fines session shows about the workflow.} Three observations are likely to transfer to other logs. First, a large share of a log may need no model: three exact-variant groups cover 44.6\% of the traces before any topic is fitted, so the list of common variants is worth checking first. Second, the group-formation mechanisms complement each other. Topics produced all groups of rounds B to E. When they stopped yielding new groups, clustering over the same topic weights found three more in round F, with lower purity (72--84\%). Third, the dominant-outcome threshold had to be lowered from 0.90 in rounds C and D to 0.75 in round F, since coherence declines once the easy structure has been carved away. Coverage therefore depends on the analyst's threshold policy as well as on the log.

\begin{figure}[t]
\centering
\includegraphics[width=\columnwidth]{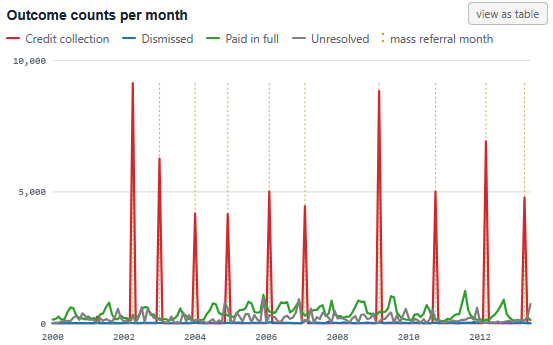}
\caption{Outcome counts per month in the Road Traffic Fines log, with the ten mass-referral months marked by dotted lines (from the temporal-dynamics view in the online supplement).}
\label{fig:drift-outcomes}
\end{figure}

\textbf{Provenance exposes process drift.} The exported contents of findings repository record which traces belong to which group, so it can be queried after the session without any new modeling. We used it to examine the end dates of the traces in each Fines group. The log itself is strongly periodic: credit-collection outcomes concentrate in ten mass-referral months (Figure~\ref{fig:drift-outcomes}). We therefore compare the groups at the level of years, which smooths over these monthly spikes. No encoding used in the session contained calendar time, so temporal structure in the groups is not an artifact of how they were formed. The groups differ markedly. Three groups have 88--94\% of their traces ending in 2000--2004, two others have 86--97\% ending in 2009--2013, and the two largest groups, which were committed first, are spread evenly over the whole span. The outcome label does not predict the pattern: the four groups dominated by \emph{unresolved} cases show four different temporal profiles. Groups defined by behavior alone thus occupy different periods of the log, which indicates that the way cases were handled changed over time. This is process drift. We had not noticed it during the session and found it afterwards from the record alone. A late concentration of open cases may also reflect right-censoring at the end of the log, so we regard these findings as pointers for a dedicated drift analysis. The online supplement gives the details.

\textbf{Computational cost.} Every step of the workflow, candidate-encoding scoring, an NMF fit at a single $K$, a full $K$-sweep, a UMAP embedding, and $k$-means clustering, completes in well under a minute, even on Road Traffic Fines, the largest of the three logs at 150{,}370 traces. With the two heaviest steps moved off the main thread (Section~\ref{sec:system}), the interactive latency an analyst experiences is dominated by their own inspection of each result, not by computation. We therefore do not report microbenchmarks. The relevant comparison is not one computation against another, but computation time against the time a human expert needs to judge, describe, and validate a candidate group, and by that comparison computation is not the bottleneck at any scale we tested.

\textbf{The embedding is a visual aid.} The UMAP embedding and the clusters help the analyst to find candidate groups. They do not drive the analysis. $k$-means works on the full topic-weight vectors $w_i \in \mathbb{R}^K$ and never uses the two-dimensional points. The embedding shows the same vectors in a form that the analyst can inspect and brush. A selected region is scored from the full vectors of its traces, so $S(G)$ and $O(G)$ do not depend on the projection. The embedding could be replaced or removed without changing the meaning or the score of any group, and the analyst decides which groups are committed.

\textbf{Limitations and threats to validity.} The graph view extends past a classical directly-follows graph for only one token family, a dashed edge per salient $n$-gram sequence; the gap, timing, and attribute-augmented families of Table~\ref{tab:encodings} have no mark on the graph and are read through the fingerprint and coherence scores instead, even when the scoring mechanism favors them, as it does for round B of the Fines session. The workflow also has no automatic stopping rule: whether a working set is homogeneous enough to stop, and whether a candidate's coherence is worth committing, are judgments left to the analyst (T1, T5) by design, so the result is not reproducible without a human in the loop.

UMAP and $k$-means are both seed- and parameter-sensitive; all three sessions fixed a seed and swept only the parameters the system exposes rather than searching the full parameter or reseeding space, so a different seed could plausibly surface a somewhat different partition, and we do not report a reseed stability sweep, since the claim under test is that the analyst can reach and record a coherent partition, not that it is the uniquely optimal one. The encoding suggestions come from a short, fixed list of candidates ranked at a fixed $K{=}4$, regardless of the $K$ eventually chosen for the working set; we have not verified that this ranking stays representative across the analyst's later choice of $K$. %

The purity and size thresholds used to decide whether a candidate was coherent enough to commit were set by inspection within each session, not fixed in advance or held out on independent data; BPI2013 in particular needed its dominant-outcome-share filter relaxed from 0.80 to 0.75 before its clustering round surfaced any qualifying group, so the resulting coverage is better read as a property of that session's threshold policy than as a fixed property of the log, and a different analyst could reasonably report a different figure for the same log. Because each committed group's traces are removed from \emph{remaining} before the next round, a different commit order could plausibly produce a somewhat different set of findings.

We do not test formally whether a group's coherence reflects a behavioral regime or merely tracks when its cases occurred. The temporal check above shows that groups differ in time, but separating behavior from calendar effects, for example a policy change, requires a concept-drift analysis of the kind developed for event logs~\cite{bose2014drift,sato2022driftsurvey}, which we leave open. Finally, our evaluation consists of usage scenarios (Section~\ref{sec:introduction}): it shows that the workflow finds interpretable, outcome-relevant sub-populations across three domains while the analyst's knowledge of the log grows traceably, not that these sub-populations are better process models than those of another method.

\section{Conclusion} \label{sec:conclusion}

Real processes rarely behave in one way, yet event logs are usually described by one model, one clustering, or one classifier. We presented TracePrism, which replaces this one-shot step with a loop that the analyst controls. In each round, the analyst picks a working set from the traces that no accepted group covers yet, and an encoding for it, guided by suggestions from the system and by knowledge of the process. The analyst then forms a candidate group, accepts it if its members are alike in process structure and in outcome, and continues with the remaining traces. A group can be reopened later under a different working set or encoding.

A recent Dagstuhl seminar 25152 on multi-faceted visual process mining identified the fixity of the trace representation as a limiting factor of current tooling. TracePrism lifts this limitation. The representation is no longer fixed once for the whole analysis. It is chosen anew for every working set, and each choice is kept on record. In the Fines session, the analyst used timing families for a working set in which activities carry no signal, switched to activities and 2-grams for the heterogeneous remainder, and added 3-grams with timing families once the obvious 2-gram patterns had been committed. Each change followed what the analyst knew or had just learned about the log, and a single fixed representation could not have supported all three steps.

The implementation and its application to three event logs show that the knowledge gained during an analysis can be documented as it grows. The resulting record is transparent and interpretable: each group is described by its structure, outcome, and attributes, and a revision is kept next to the decision it replaces. An analysis of an event log can then leave behind more than a final model: a record of how it was reached, which others can audit, reopen, and build on.

\section*{Acknowledgments}
This work was done with support of the Lamarr Institute for Machine Learning and Artificial Intelligence. This paper is a result of the discussions at the Dagstuhl Seminar 25152, “Multi-Faceted Visual Process Mining and Analytics”.

\balance
\bibliographystyle{IEEEtran}
\bibliography{references}

\end{document}